\documentclass[aps,prl,reprint,superscriptaddress,floatfix
]{revtex4-1}
\usepackage{graphicx}
\usepackage{color}
\usepackage{dcolumn}
\usepackage{bm}
\usepackage{amssymb}
\usepackage{color,amsmath}
\usepackage{soul,xcolor} 
\setstcolor{red}
\usepackage{siunitx}
\usepackage{epstopdf}
\usepackage{gensymb}

\DeclareSIUnit\angstrom{\text {Å}}

\usepackage{siunitx}
\begin{document}
\title{Direct magnetic imaging of 
fractional Chern insulators in twisted MoTe$_2$ with a superconducting sensor}

\author{Evgeny Redekop}
\affiliation{Department of Physics, University of California, Santa Barbara, CA 93106, USA}
\author{Canxun Zhang}
\affiliation{Department of Physics, University of California, Santa Barbara, CA 93106, USA}
\author{Heonjoon Park}
\affiliation{Department of Physics, University of Washington, Seattle, WA 98195, USA}
\author{Jiaqi Cai}
\affiliation{Department of Physics, University of Washington, Seattle, WA 98195, USA}
\author{Eric Anderson}
\affiliation{Department of Physics, University of Washington, Seattle, WA 98195, USA}
\author{Owen Sheekey}
\affiliation{Department of Physics, University of California, Santa Barbara, CA 93106, USA}
\author{Trevor Arp}
\affiliation{Department of Physics, University of California, Santa Barbara, CA 93106, USA}
\author{Grigory Babikyan}
\affiliation{Department of Physics, University of California, Santa Barbara, CA 93106, USA}
\author{Samuel Salters}
\affiliation{Department of Physics, University of California, Santa Barbara, CA 93106, USA}
\author{Kenji Watanabe}
\affiliation{Research Center for Electronic and Optical Materials, National Institute for Materials Science, 1-1 Namiki, Tsukuba 305-0044, Japan}
\author{Takashi Taniguchi}
\affiliation{Research Center for Materials Nanoarchitectonics, National Institute for Materials Science, 1-1 Namiki, Tsukuba 305-0044, Japan}
\author{Xiaodong Xu}
\affiliation{Department of Physics, University of Washington, Seattle, WA 98195, USA}
\affiliation{Department of Materials Science and Engineering, University of Washington, Seattle, WA 98195, USA}
\author{Andrea F. Young}
\email{andrea@physics.ucsb.edu}
\affiliation{Department of Physics, University of California, Santa Barbara, CA 93106, USA}
\begin{abstract}
\end{abstract}

\maketitle

\textbf{
In the absence of time reversal symmetry, orbital magnetization provides a sensitive probe of topology and interactions, with particularly rich phenomenology in Chern insulators where  topological edge states carry large equilibrium currents.  
Here, we use a nanoscale superconducting sensor \cite{finkler_self-aligned_2010} to map the magnetic fringe fields in twisted bilayers of MoTe$_2$, where transport \cite{park_observation_2023, xu_observation_2023} and optical sensing \cite{cai_signatures_2023,zeng_thermodynamic_2023} experiments have revealed the formation of fractional Chern insulator (FCI) states at zero magnetic field. 
At a temperature of $\SI{1.6}{K}$, we observe oscillations in the local magnetic field associated with fillings $\nu=-1,-2/3,-3/5,-4/7$ and $-5/9$ of the first moir\'e hole band, consistent with the formation of FCIs at these fillings. 
By quantitatively reconstructing the magnetization, we determine the local thermodynamic gaps of the most robust FCI state at $\nu=-2/3$, finding $^{-2/3}\Delta$ as large as $\SI{7}{meV}$.  
Spatial mapping of the charge density- and displacement field-tuned magnetic phase diagram further allows us to characterize sample disorder, which we find to be dominated by both inhomogeneity in the effective unit cell area\cite{uri_mapping_2020} as well as inhomogeneity in the band edge offset and bound dipole moment.   
Our results highlight both the challenges posed by structural disorder in the study of twisted homobilayer  moir\'e systems and the opportunities afforded by the remarkably robust nature of the underlying correlated topological states.}

\begin{figure*}[ht!]
\includegraphics[width=6.5in]{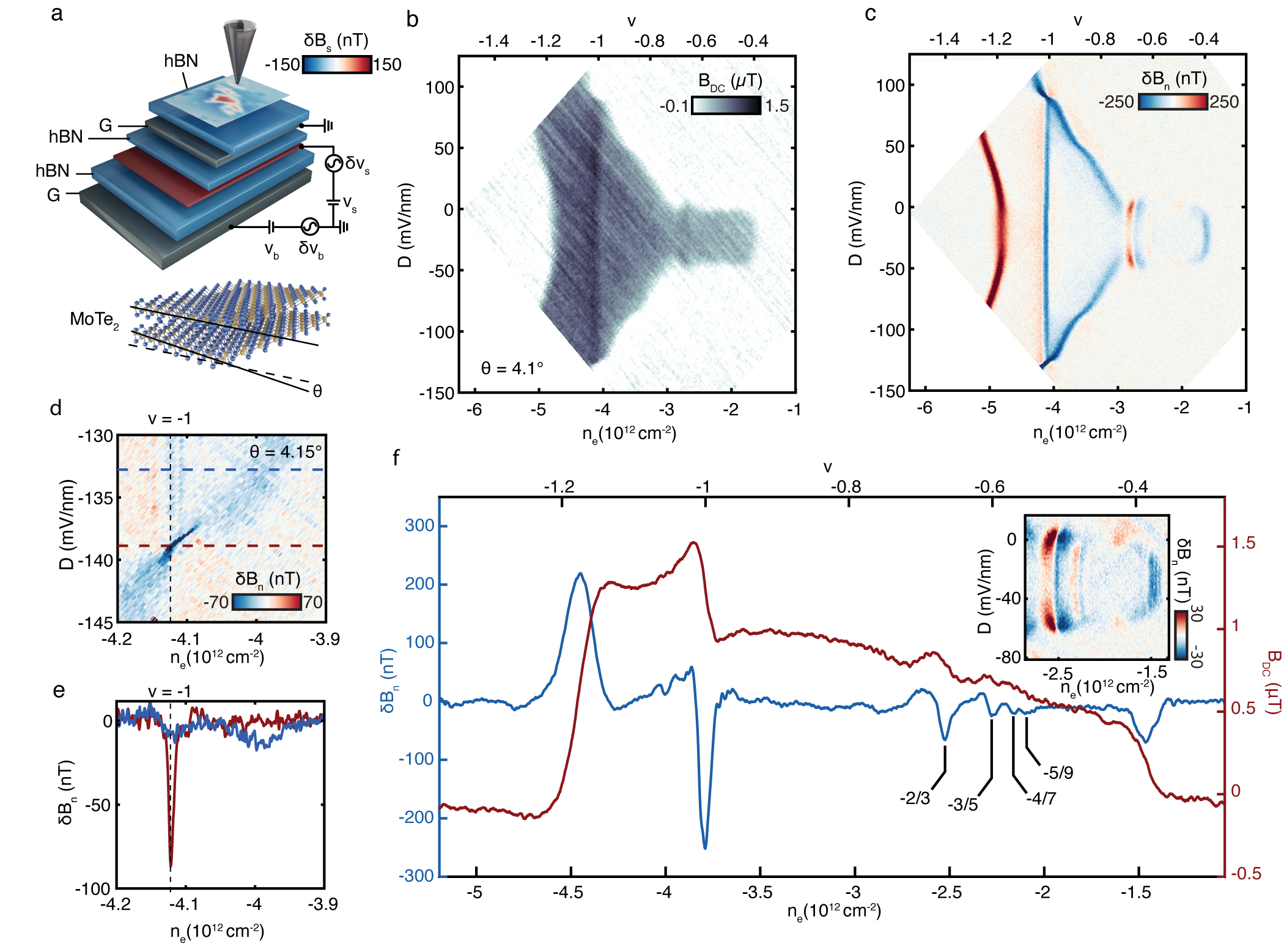}
\caption{\textbf{Local magnetometry of FCIs in twisted MoTe$_2$.}
(\textbf{a}) Schematic of the experimental geometry, showing a nSOT sensor above a van der Waals heterostructure consisting of a dual graphite-gated MoTe$_2$ bilayer with nominal $\theta=3.9^\circ$ misalignment angle. 
Static voltages $V_s$ and $V_b$ and modulated voltages $\delta V_s$ and $\delta V_b$ are applied to the sample and bottom gate as shown, allowing finite-frequency readout of the modulated magnetization, shown in overlay for finite $\delta V_s$ and $\delta V_b=0$ at moir\'e filling $\nu=-1$ in Sample B.  
(\textbf{b}) 
Static magnetic field $B_\mathrm{DC}$ measured at a single point as a function of charge carrier density $n_e$ and applied electric displacement field $D$ at \SI{1.6}{K} in Sample A.  
(\textbf{c}) $\delta B_n$ with an applied modulation of $\delta n\approx \SI{1.5d10}{cm^{-2}}$, measured at the same position as data in panel \textbf{b}.  
(\textbf{d}) $\delta B_n$ measured with $\delta n\approx \SI{1.5d9}{cm^{-2}}$ near the valley polarization transition at $\nu=-1$. The sharp signal is indicative of a first-order phase transition where the magnetization vanishes. 
(\textbf{e}) Two traces extracted from panel \textbf{d}. 
(\textbf{f}) 
$\delta B_n$ and $B_\mathrm{DC}$ measured at $T=\SI{1.6}{K}$, $B = \SI{34}{mT}$ and $D = \SI{-30}{mV/nm}$; for $\delta B_n$ the applied $\delta n\approx \SI{1.2d10}{cm^{-2}}$. 
We identify the peak at $\nu\approx -1.175$ and dip at $\nu\approx-0.39$ with the boundaries of the magnetic phase, and the five minima at $\nu=-1,-2/3,-3/5,-4/7$, and $-5/9$ with the edge state magnetization of Chern insulator states. 
Inset: $\delta B_n$ measured with an applied modulation $\delta n\approx \SI{1.5d9}{cm^{-2}}$ in the FCI regime, showing $D$-independent minima at $\nu=-2/3$ and $-3/5$ within the valley-polarized regime.
}
\label{fig1}
\end{figure*}


\subsubsection*{\textbf{Background}} 

Fractional Chern insulators (FCIs) are generalizations of the fractional quantum Hall states to lattice systems with broken time reversal symmetry. 
Interest in FCIs arises from the fact that they may emerge as the ground state of interacting fermions on a lattice, including at zero magnetic field where time reversal symmetry is broken spontaneously.  
Because FCIs are driven by electron--electron interactions, the energy gap $\Delta$ separating the incompressible ground state from its charged excitations is expected to scale with interparticle separation.  In a lattice system, this is given by the inverse of the lattice constant $\lambda$, $\Delta\propto e^2/\lambda$ (where $e$ denotes the electron charge). For sufficiently small $\lambda$, these energy gaps may be comparable to or larger than the energy gaps of fractional quantum Hall states in partially filled Landau levels at experimentally accessible magnetic fields, where the interparticle spacing is set by the magnetic length, $\ell_B\approx \SI{25.7}{nm}/\sqrt{B \mathrm{[T]}}$.

FCIs were first observed experimentally in moir\'e heterostructures composed of graphene and hexagonal boron nitride (hBN) at partial fillings of topological Harper-Hofstadter bands \cite{spanton_observation_2018} at \SI{30}{T} magnetic fields; in this context the applied magnetic field plays a key role in forming the bands leading to energy gaps comparable to those observed in conventional fractional quantum Hall states. Prospects for realizing more robust lattice-based Chern insulators improved with the realization of narrow, topologically nontrivial bands at zero magnetic field in moir\'e systems including twisted bilayer graphene \cite{cao_correlated_2018}, rhombohedral graphene multilayers aligned to hBN \cite{chen_evidence_2019}, and transition metal dichalcogenide bilayers \cite{regan_mott_2020,tang_simulation_2020,wang_correlated_2020,shimazaki_strongly_2020,xu_correlated_2020}. 
In these systems, experiments showed evidence for time reversal symmetry breaking via spontaneous valley polarization, manifesting most dramatically with the observation of quantized integer anomalous Hall effects at zero magnetic field \cite{
sharpe_emergent_2019, serlin_intrinsic_2020, chen_tunable_2020,li_quantum_2021,anderson_programming_2023}, as 
well as FCI states at finite applied magnetic fields \cite{xie_fractional_2021}. 
Most recently, zero magnetic field FCIs were discovered in rotationally faulted bilayers of MoTe$_2$ near 4$^\circ$ angle \cite{cai_signatures_2023,zeng_thermodynamic_2023,park_observation_2023,xu_observation_2023, ji_local_2024} and in rhombohedral graphene pentalayers aligned to a hBN substrate \cite{lu_fractional_2023}. 
These observations raise key questions about the microscopic origin of these states, their competition with other correlated ground states, and the possibility for realizing novel topological phases absent in partially filled Landau levels, leading to a growing body of theoretical work \cite{dong_composite_2023,wang_fractional_2024,yu_fractional_2024,morales-duran_magic_2023, jia_moire_2023, sheng_quantum_2024, dong_anomalous_2023, soejima_anomalous_2024, goldman_zero-field_2023}. 

\subsubsection*{\textbf{Local magnetometry of fractional Chern insulators}}
Here we take advantage of the compatibility of zero-magnetic field FCIs with superconducting sensors to perform ultrasensitive magnetometry of the fringe magnetic fields associated with spin and orbital magnetism in twisted MoTe$_2$ on the submicron scale.  
In a Chern insulator, equilibrium currents carried by the chiral edge states contribute a topological magnetization  
\begin{equation}
\Delta m=\frac{C\Delta}{\Phi_0},
\label{eq:gap}
\end{equation} 
which represents the change in magnetization across the incompressible bulk gap (here $C$ is the total Chern number, $\Delta$ is the thermodynamic energy gap, and $\Phi_0$ is the non-superconducting flux quantum).  
This universal contribution occurs atop a non-universal background of spin and orbital magnetic moments arising from the filled electronic states in the sample bulk.  
Previous experiments have used nanoscale superconducting quantum interference device on tip (nSOT) \cite{vasyukov_scanning_2013,anahory_squid--tip_2020} to detect orbital currents associated with the formation of integer quantum Hall states in monolayer graphene \cite{uri_mapping_2020}, quantum anomalous Hall states in both graphene and dichalcogenide moir\'e systems \cite{tschirhart_imaging_2021,tschirhart_intrinsic_2023}, 
and the non-topological orbital magnetization associated with orbitally polarized metallic states in both moir\'e and non-moir\'e graphene systems \cite{grover_imaging_2022, arp_intervalley_2023}. However, magnetic imaging of fractionalized phases has not been reported. 

We use an indium nSOT sensor with effective diameter of approximately $\SI{200}{nm}$ to map the fringe magnetic field above several twisted bilayer MoTe$_2$ samples. 
Our sensors represent a significant improvement over the state-of-the-art \cite{anahory_squid--tip_2020}, with a measured sensitivity of $\SI{0.2}{nT/\sqrt{Hz}}$ (see Extended Data Fig. \ref{SQUIDsensitivity} and Methods) corresponding to $\SI{4}{n\Phi_0/\sqrt{Hz}}$, where $\Phi_0$ is the superconducting flux quantum.
Our samples consist of a twisted bilayer MoTe$_2$ active layer encapsulated by hBN dielectrics with graphite top and bottom gates (see Fig. \ref{fig1}a and Extended Data Fig. \ref{Devices}).  Together, the gates allow independent control of the charge carrier density $n_e$ and electric displacement field $D$. 
The graphite gates are transparent to the fringe magnetic fields, making magnetic imaging an ideal tool for probing the $n_e$ and $D$-tuned phase diagram.  We measure the fringe fields either in direct current mode via the static magnetic field $B_\mathrm{DC}$ at fixed $n$ and $D$ or in alternating current mode through lock-in readout of the magnetic response to modulations applied to the sample and/or top and bottom gate voltages.
Appropriate choice of the relative magnitude and phase of these voltages allows us to measure either the density- or displacement field derivatives of the fringe field, $\delta B_n\approx \frac{\partial B_\mathrm{DC}}{\partial n_e}\delta n_e$ and $\delta B_D\approx \frac{\partial B_\mathrm{DC}}{\partial D}\delta D$.

\begin{figure*}[ht!]
\includegraphics[width=6.5in]{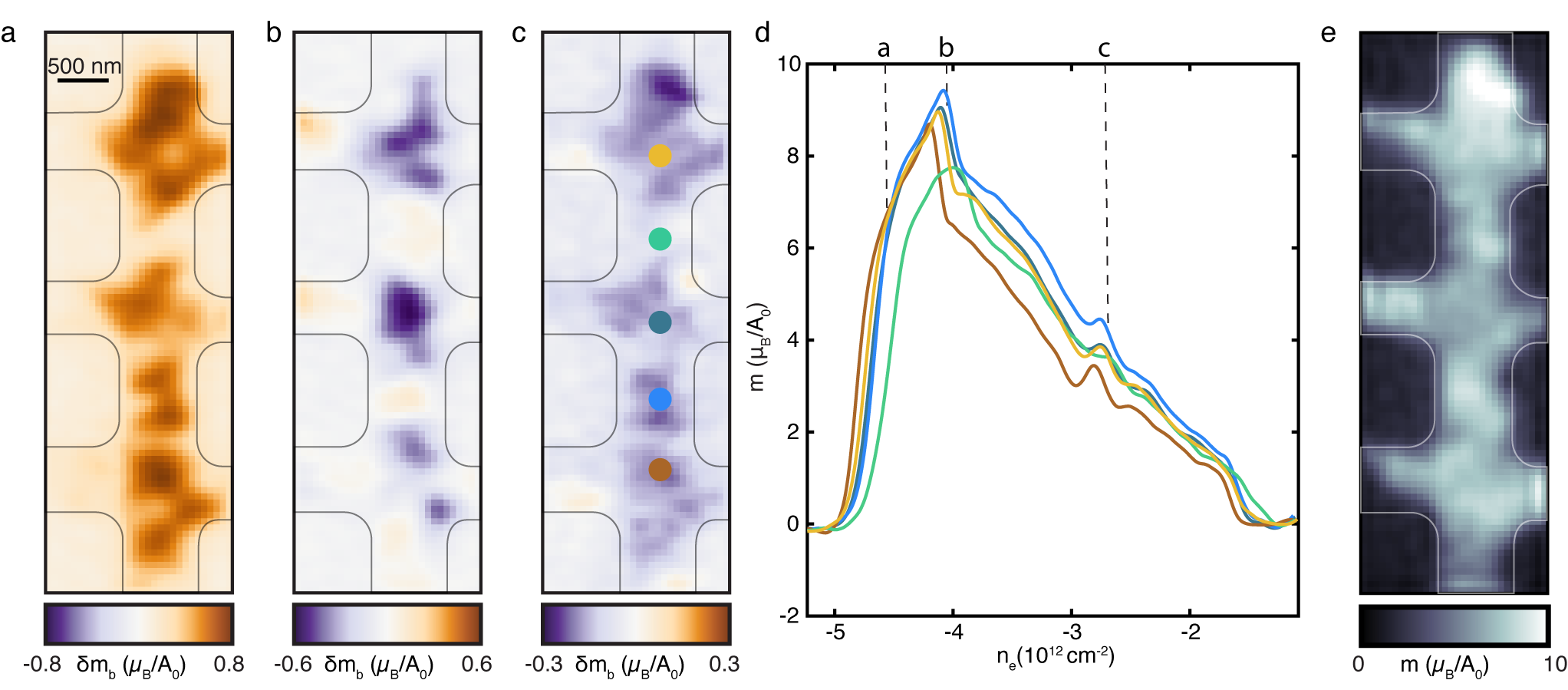}
\caption{\textbf{Reconstructing local magnetization.}
(\textbf{a}) Differential magnetization $\delta m_b$ reconstructed from spatial maps of $\delta B_b\approx (\partial B/\partial V_b)\delta V_b$, measured at $D = \SI{-30}{mV/nm}$ and density corresponding to  $\nu \approx -1.2$, 
(\textbf{b}) $\nu \approx -1$, and 
(\textbf{c}) $\nu \approx -0.66$. Here $A_0=\SI{23.7}{nm^2}$ is the superlattice unit cell area corresponding to a twist angle of \SI{3.89}{\degree}.
(\textbf{d}) Total out-of-plane magnetization $m$, 
calculated by integrating the differential magnetization for the spatial positions indicated in panel \textbf{c}. 
(\textbf{e}) Spatial image of $m$ at $D = \SI{-30}{mV/nm}$ and $\nu = -1.12$.}
\label{fig2}
\end{figure*}

Figure \ref{fig1}b shows $B_\mathrm{DC}$ measured locally at a height $\SI{50}{nm}$ above the surface of sample A. Magnetic signal is observed in a broad region of the phase diagram centered at low displacement field, consistent with previous optical and transport measurements \cite{anderson_programming_2023, park_observation_2023,zeng_thermodynamic_2023}. We associate this regime with a valley-imbalanced ferromagnetic phase. Features previously identified as Chern insulators at $\nu=-1$ and $\nu=-2/3$ are visible in this image as subtle vertical lines. These become plainly visible in measurements of $\delta B_n$, shown in Fig. \ref{fig1}c.  In this contrast mode, the \textit{decrease} in magnetization with \textit{increasing} chemical potential, characteristic of the edge states of Chern insulators with negative $C$, manifests as a sharp, negative (blue) signal that appears at a $D$-independent density. Similar phase diagrams were obtained from three samples, as shown in Extended Data Fig. \ref{Devices}. The charge carrier density in the tMoTe$_2$ may be independently calibrated using magnetic features associated with Landau levels in the graphite top gate, and measurements in Device C confirm that the negative $\delta B_n$ feature associated with the $\nu=-2/3$ state occurs at filling factor $\nu=0.66\pm.02$ (see Extended Data Fig. \ref{LLs}).

Our high sensitivity local measurement allows us to examine several aspects of the phase diagram that have been ambiguous in previous studies of the same system.  
For example, we find that, while the signal associated with the Chern insulator gap is the strongest at zero effective displacement field (see Extended Data Fig. \ref{Ch_displacement_field}), it remains finite until the sharp, $D$-tuned phase transition to a non-magnetic phase. This is confirmed by low excitation, high resolution measurements of $\delta B_n$ in the vicinity of the phase transition at large negative $D$ shown in Fig. \ref{fig1}d, where an exceptionally sharp feature associated with a decrease in magnetization appears at $\nu=-1$. As shown in Fig. \ref{fig1}e, the fringe magnetic field associated with this feature is about one order of magnitude larger than those associated with the Chern insulator edge states at $\nu=-1$, and the large negative signal occurs over a narrow range of $D$ and $n_e$. 
This is expected for a first order phase transition between a valley-imbalanced Chern insulator and a topologically trivial insulator with no net magnetization (see additional data in Extended Data Fig. \ref{AbsenceOfTrivialMagnetization}). We conclude that the first order phase transition in valley polarization occurs before any displacement field-tuned change in band topology, and find no evidence for a valley-imbalanced metallic state. Notably, while the valley-imbalanced transition is sharp at $\nu=-1$, where the transition occurs between a Chern insulator and topologically trivial correlated insulator\cite{park_observation_2023}, it is broad in the metallic regimes between commensurate filling factors.

Figure \ref{fig1}f shows a high resolution trace of both $B_{DC}$ and $\delta B_n$ acquired at $T=\SI{1.6}{K}$, $B=\SI{34}{mT}$ and $D=\SI{-30}{mV/nm}$. 
In this dataset, we observe oscillations in the local magnetic field, with oscillation minima associated with band fillings $\nu=-1,-2/3,-3/5,-4/7$ and $-5/9$. High resolution data acquired near filling $\nu=-1/2$ as a function of both $n_e$ and $D$ (Fig. \ref{fig1}f, inset) shows negative $\delta B_n$ features at $\nu=-2/3$ and $\nu=-3/5$, which (mimicking the behavior at $\nu=-1$) persist to the valley polarization transition. We associate the negative features with the FCI gaps; notably, however, they are accompanied by a positive $\delta B_n$ feature at slightly higher hole density. We associate this feature with a non-topological magnetization that arises for small doping of the system away from the FCI gaps. 

Current theoretical understanding of twisted bilayer MoTe$_2$ suggests that the single-particle wave functions of the lowest energy moir\'e hole band in a single valley resemble those of the lowest energy Landau level in a two-dimensional electron system, with an emergent composite Fermi liquid state at half filling \cite{dong_composite_2023,goldman_zero-field_2023}.  As in fractional quantum Hall systems in partially filled Landau levels, then, a sequence of incompressible states are observed at fillings $p/(2p\pm1)$ corresponding to the quantum oscillations of the composite fermions in the effective magnetic field \cite{jain_composite-fermion_1989}, which in this case is completely interaction induced property of the magnetic ground state. 
In this picture, the oscillations we observe in the \textit{physical} fringe magnetic field as a function of the density may be associated with de Haas-van Alphen oscillations of the composite fermions in the \textit{emergent} magnetic field.  Our observation of oscillations in the magnetization, $m=dF/dB$, can be taken as direct evidence for the emergence of topological gaps whose charge density is magnetic field dependent.

\begin{figure*}[ht!]
\includegraphics[width=6.5in]{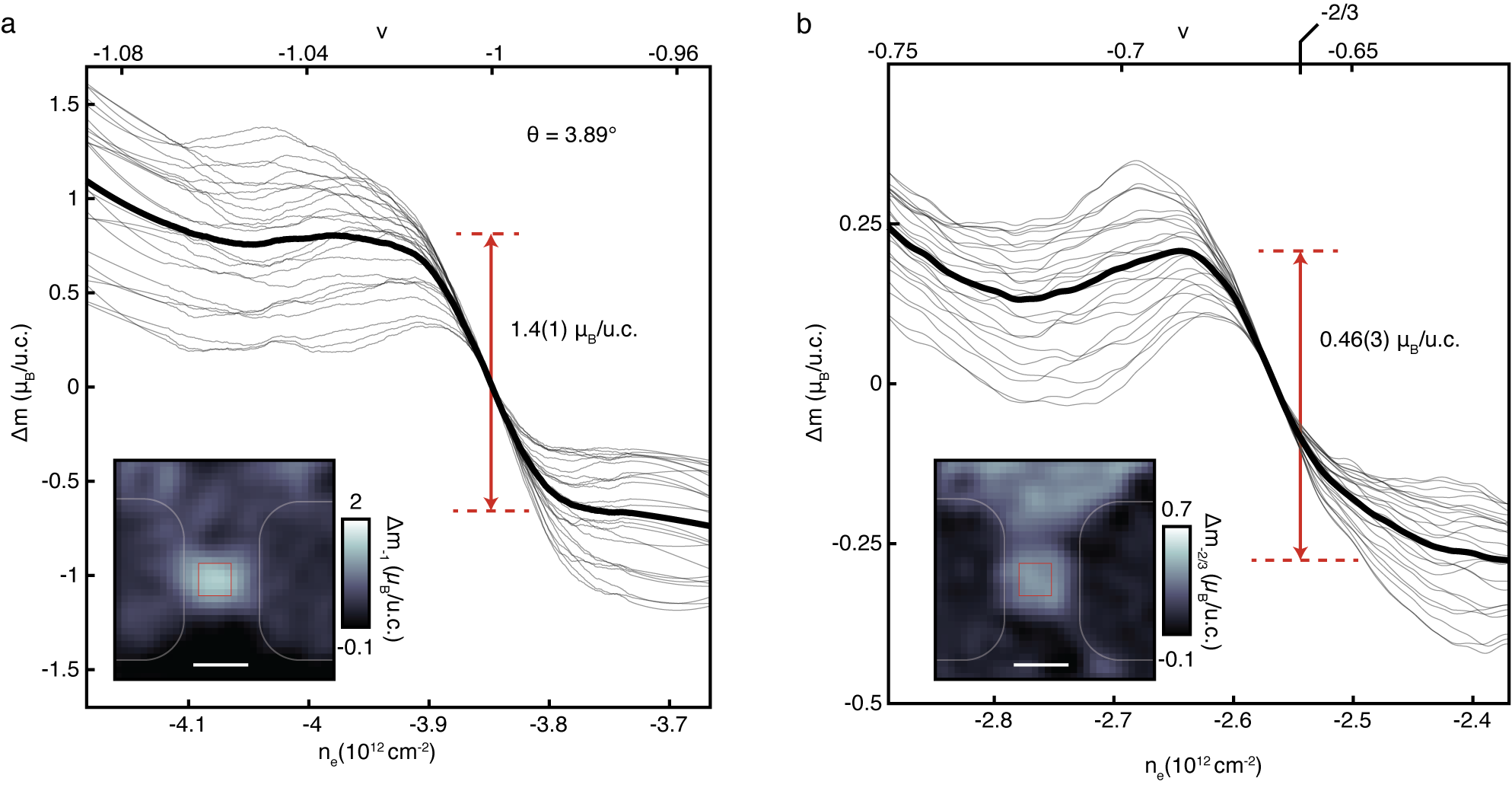}
\caption{\textbf{Thermodynamic gaps.}
(\textbf{a}) Magnetization change $\Delta m$ (referenced to the center of the $\nu=-1$ gap) in a region where the local twist angle $\theta\approx 3.89\degree$. The inset shows a spatial scan of $\Delta m_{-1}$ the difference in magnetization on the boundaries of the Chern insulator at $\nu=-1$.
We find a change of $\Delta m_{-1}=\SI{1.4\pm0.1}{\mu_B}$/u.c. across $\nu=-1$, corresponding to $^{-1}\Delta=\SI{14\pm1}{meV}$ gap for $C=-1$.
(\textbf{b}) $m$ for $\nu = -2/3$ (referenced to the center of the $\nu=-2/3$ gap), measured in the same grid points. Inset shows the difference in magnetization on the boundaries of the Chern insulator at $\nu=-2/3$. $\Delta m_{-2/3}=\SI{0.46\pm0.03}{\mu_B}$/u.c., corresponding to thermodynamic gap $^{-1}\Delta=\SI{7\pm0.5}{meV}$, assuming $C=-2/3$.
}
\label{fig3}
\end{figure*}

\subsubsection*{\textbf{Quantifying local magnetization}}

The data shown in Fig. \ref{fig1} provides a qualitative picture of the microscopic phase diagram.  To quantitatively measure the magnetization,  we take spatial scans of $B_\mathrm{DC}$ and $\delta B_b$ and use an inversion algorithm to reconstruct the static magnetization $m$ and differential magnetization change in response to a change in bottom gate voltage, $\delta m_b=\partial m/\partial V_b \times \delta V_b$ (see Methods and Extended Data Fig. \ref{MRecon}). 
This analysis assumes that magnetic moments point only in the out-of-plane direction, an assumption that is well justified in strongly spin-orbit coupled twisted MoTe$_2$.  
Figures \ref{fig2}a, b and c show $\delta m_b$ images acquired at applied $D=\SI{-30}{mV/nm}$ and $n_e$ corresponding to $\nu \approx -1.1$, $\nu \approx -1$ and $\nu \approx -2/3$ respectively. 
The sample magnetization is inhomogenous on submicron scales, a feature universal to all samples measured. 
Figure \ref{fig2}d shows the $n_e$-dependent total $m$, obtained by integrating $\delta m_b$, at several different points in the device shown in Fig.\ref{fig2}c. 
These data are consistent with a large, non-topological orbital magnetization in the ferromagnetic phase ranging from 6 to 8 Bohr magnetons ($\mu_B$) per hole. 
This agrees with estimates for the renormalization of the spin moment arising from atomic scale spin-orbit coupling determined from band theory and optical experiments \cite{deilmann_ab_2020,wozniak_exciton_2020,robert_measurement_2021}.
Despite the strong spatial inhomogeneity in the internal structure of the ferromagnetic region, the entire sample is magnetized, consistent with valley polarization being a robust feature of the phase diagram across a range of sample parameters (Fig. \ref{fig2}e).

Our quantitative reconstruction of the magnetization may also be used to directly determine the Chern insulator gaps via Eq. \eqref{eq:gap}. Figures \ref{fig3}a-b show $m$ in the vicinity of $\nu=-1$ and $\nu=-2/3$. 
We extract the change in magnetization from the extent of the negative-slope regions at each rational filling factor, obtaining $\Delta m_{-1}=\SI{1.4\pm0.1}{\mu_B}$/u.c. and $\Delta m_{-2/3}=\SI{0.46\pm0.03}{\mu_B}$/u.c.
Assuming $C=\nu$ (consistent with transport measurements of the same sample \cite{park_observation_2023}), 
this yields thermodynamic energy gaps of $^{-1}\Delta=\SI{14\pm1}{meV}$ and $^{-2/3}\Delta=\SI{7\pm0.5}{meV}$.  


\begin{figure*}[ht!]
\includegraphics[width=7in]{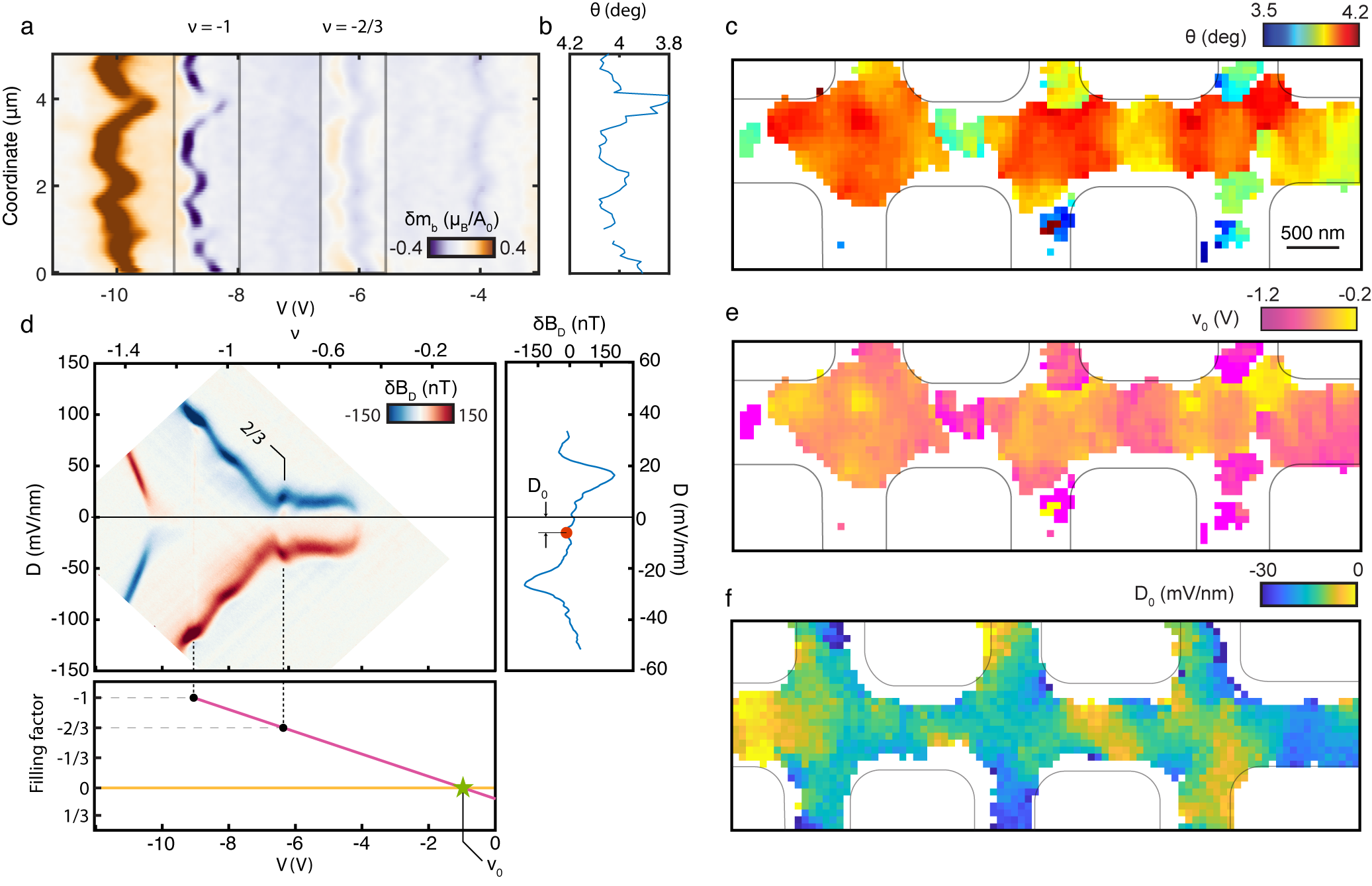}
\caption{\textbf{Sources of disorder.}
(\textbf{a}) $\delta m_b$ as a function $V\equiv V_T+V_B$ along a trajectory running along the center of the device. 
Features near $V=\SI{-9}{V}$ and $\SI{-6}{V}$ correspond to topological gaps at $\nu = -1$ and $\nu = -2/3$, respectively. 
(\textbf{b}) Effective interlayer twist angle along the spatial trajectory of panel \textbf{a}, extracted from the density difference between $\nu=-1$ and $\nu=-2/3$ magnetization features. 
(\textbf{c}) Local twist angle map in device A. 
(\textbf{d}) $\delta B_{D}$ as a function of $V$ and $D$.  
In addition to the twist angle, the voltages corresponding to Chern insulators at $\nu = -1$ and $\nu = -2/3$ are used to extract the valence band offset, $V_0$, corresponding to $n_e=0$. The $D$-induced valley-polarization transitions, meanwhile, are used to extract the displacement field offset $D_0$. 
(\textbf{e}) Spatial map of$V_0$ in device A. 
(\textbf{f}) Spatial map  $D_0$ in device A. 
}
\label{fig4}
\end{figure*}
The ranges in measured gap sizes reflect the spread in measured values in the small area outlined in red in Figs. \ref{fig3}a-b, rather than experimental uncertainty.  
Our determination of the gap sizes is, however, susceptible to several sources of error. 
While some uncertainty is contributed by the calibration of our magnetometer (which we measure to a reproducibility of approximately $5\%$), systematic errors are likely more significant. One source of error arises from the fact that our measurement window does not capture the entire region of the physical magnetic field.  
Because magnetic dipole fields are long-range, the magnetic inversion algorithm takes as an input both the field in the measured region and an assumed field outside the measured region; in the data shown in the main text, this field is assumed to be zero.  
To estimate the magnitude of the error resulting from this assumption, we compare $m$ calculated using `zero-padding' to calculations done with different physically motivated  padding assumptions (see methods and Extended Data Fig. \ref{OutsideSignalGapReconstruction}). Across several models, we find that $m$ is consistent to within 10\%. A second source of uncertainty is conceptual, and arises from our empirical definition of the thermodynamic gap, where we assume that the region of steepest negative slope in $m$ arises entirely from the chiral edge states. While this is known to be true in the clean limit for fractional Hall states in partially filled Landau levels, theoretical calculations accounting for the effects of finite disorder and inhomogenous Berry curvature are not available to justify this empirical definition of the thermodynamic energy gap. 

In a clean quantum Hall system, the thermodynamic gap measures the energy to add one electron of charge to the gapped ground state. In states whose excitations carry only integer charge, this is equivalent to the thermal activation gap.  
In states whose elementary excitations carry fractional charge $e^*=e/q$, however, the thermal activation gap measured in transport is expected to be smaller than the thermodynamic gap by a factor of $q$. 
In a disorder-free, partially filled Landau level with Coulomb interactions, the ratio of the thermodynamic gaps at $\nu=1$ and $\nu=1/3$ is expected to be approximately 4\cite{morf_monte_1986}; we find a ratio of approximately 2--i.e., the fractional state is larger in comparison to the integer state than for a Landau level system. 
Our data also show a notable when compared with transport measurements.  From our measurements, $^{-1}\Delta$ corresponds to a thermal activation gap of $\SI{150}{K}$, approximately five times larger than the $\SI{30}{K}$ found in transport measurements\cite{park_observation_2023}.  
However, our measurement of $^{-2/3}\Delta\approx\SI{7}{meV}$ corresponds to an estimated thermal activation gap of approximately $\SI{27}{K}$, roughly\textit{consistent} with transport measurements\cite{park_observation_2023}. Quantitatively reconciling these observations requires a detailed theory that accounts for both band effects unique to lattice Chern bands as well as the contrasting effects of disorder on the thermodynamic and transport gaps at integer and fractional filling.

As a final point of comparison, the energy scales measured here are also similar in scale to thermodynamic gaps measured in monolayer graphene fractional quantum Hall systems at $B\approx \SI{14}{T}$ in a similar electrostatic geometry \cite{yang_experimental_2021}.  
We note, however, that sample inhomogeneity on length scales smaller than that of our local probe may well contribute to the lowering of the measured thermodynamic gaps in twisted MoTe$_2$ relative to the intrinsic gap size.
The intrinsic gaps may thus be even larger than reported here, consistent with theoretical analyses that find MoTe$_2$ bilayers to be equivalent to conventional quantum Hall states at magnetic fields as large as $\SI{160}{T}$ \cite{dong_composite_2023}.  

\subsubsection*{\textbf{Sources of disorder in twisted MoTe$_2$}}

A key consideration in the interpretation of both transport and bulk thermodynamic data in moir\'e systems is sample disorder. In particular, variations in structural parameters can make the same Chern insulator state occur at different values of the applied gate voltages in different parts of the sample; the sample will then not be uniformly gapped and measureents will be strongly dependent on the size and shape of the probed region. Macroscopic thermodynamic probes average both incompressible and compressible regions, reducing the measured thermodynamic gap, while transport measurements may see edge state transport shorted by bulk conducting regions of the sample. The effect of structural disorder is evident in Fig. \ref{fig4}a, which shows $\delta m_b$ as a function of the sum of the applied gate voltages to the top and bottom gates, $V\equiv V_t+V_b$ (with $D$ held constant) and the spatial coordinate along a trajectory that runs along the `spine' of the Hall bar sample. 
The applied gate voltage required to reach the same moir\'e density varies by $\approx 10\%$. As the Chern insulator states at $\nu=-1$ and $\nu=-2/3$ occur at fixed, known filling of the moir\'e superlattice, the voltages at which they appear allow us to extract the `moir\'e density', defined as $n_M^{-1}=\frac{\sqrt{3}a^2}{2\sin(\theta)}$ via the relation $c(V_{-1}-V_{-2/3})=n_M/3$ where $c$ is the capacitance per unit area of the two gates. 
The effective interlayer twist angle $\theta$ corresponding to the trajectory in Fig. \ref{fig4}a is plotted in Fig. \ref{fig4}b, while a map of the effective twist angle throughout the device is shown in Fig. \ref{fig4}c. 
Qualitatively, the twist angle map is reminiscent of twisted bilayer graphene \cite{uri_mapping_2020} with regions of approximately uniform twist angle separated by domain walls where the effective twist angle changes suddenly. 

We also identify additional microscopic sources of disorder.  Fig. \ref{fig4}d shows a measurements of $\delta B_D$ as a function of $D$ and $V$. 
In addition to the twist angle, $V_{-1}$ and $V_{-2/3}$ also allow us to determine the threshold voltage associated with the valence band edge, $V_0=3 V_{-2/3}-2V_{-1}$.  This quantity, which we term the band edge offset, may be determined by the band gap of the MoTe$_2$ layers as well as bound electric charges (for example in an impurity band) which must be filled before the first itinerant hole populates the MoTe$_2$ valence band. 
Fig. \ref{fig4}e shows a spatial map of this quantity, which varies by as much as one volt across the sample. Variations in this parameter are confirmed by chemical potential sensing measurements that leverage the magnetic response of the top graphite gate (see Extended data Fig. \ref{LLs}). 
We also find evidence for a built-in dipole moment. 
As shown in Fig. \ref{fig4}d, the ferromagnetic region of the phase diagram is symmetric about a fixed but non-zero value of $D$.  
We associate this displacement field offset $D_0$ with a built-in electric field, which we find varies spatially on the micron scale as shown in Fig. \ref{fig4}f.  This electric field may be associated with heterostrain in the MoTe$_2$ bilayer, which breaks the layer-inversion symmetry of the idealized bilayer system at $D=\SI{0}{mV/nm}$ and induces bound dipole charge within the layers.  Despite these sources of inhomogeneity, FCI states are observed over most of the sample area, suggesting a certain degree of robustness of the underlying phenomena.  

\subsubsection*{\textbf{Conclusion}}
In conclusion, our results highlight both the promise and challenges of twisted homobilayer moir\'e materials for the study of FCI physics. On one hand, sizes of the measured FCI gaps make twisted MoTe$_2$ one of the most robust fractional quantum Hall systems studied. On the other hand, the strong spatial inhomogeneity poses a challenge for future experiments. Interferometric detection of quasiparticle statistics \cite{nakamura_direct_2020}, for example, typically requires highly uniform two-dimensional electron systems where the trajectory of the current-carrying edge states can be precisely controlled. 

\subsubsection*{\textbf{Acknowledgments}}
The authors thank Liang Fu, Taige Wang, and Michael Zaletel for discussions.
Work at UCSB was primarily supported by the Army Research Office under award W911NF-20-2-0166. 
E.R. and O.S. acknowledge support by the National Science Foundation through Enabling Quantum Leap: Convergent Accelerated Discovery Foundries for Quantum Materials Science, Engineering and Information (Q-AMASE-i) award number DMR-1906325. 
A.F.Y. acknowledges additional support by the Gordon and Betty Moore Foundation EPIQS program under award GBMF9471. 
The work at University of Washington is supported by DoE BES under award DE-SC0018171. Device fabrication used the facilities and instrumentation supported by NSF MRSEC DMR-230879. 
K.W. and T.T. acknowledge support from the Elemental Strategy Initiative conducted by the MEXT, Japan (Grant Number JPMXP0112101001) and JSPS KAKENHI (Grant Numbers 19H05790, 20H00354 and 21H05233). 

\section*{Competing interests}
The authors declare no competing interests.

\section*{Data availability}
The data that support the plots within this paper are available from the corresponding author upon reasonable request.

\bibliographystyle{apsrev4-1}
\bibliography{references}

\clearpage
\section{Methods}
\subsection{Device Fabrication}
Devices used in this study are fabricated using methods previously described in the literature. Transport data from device A was previously reported in Ref. \cite{park_observation_2023}, where it appears as device ``D($3.7\degree$)''). Device B was previously studied in Ref. \cite{cai_signatures_2023}.

\subsection{nSOT sensor fabrication and local magnetometry measurements}

We perform magnetic imaging with a superconducting quantum interference device on the apex of a sharp quartz pipette (nSOT) \cite{finkler_self-aligned_2010, vasyukov_scanning_2013, anahory_squid--tip_2020}. 
We use a quartz micropipette with inner tube diameter of $\SI{0.5}{mm}$ to pull a sharp tip with the apex diameter of $\approx\SI{150}{nm}$. To form the coarse contacts, we deposit gold films by electron beam evaporation at a deposition rate of $\SI{2}{\angstrom/s}$ to produce a thickness of ($\SI{50}{\angstrom}$ Ti/$\SI{500}{\angstrom}$ Au). A shunt resistor is further deposited with ($\SI{80}{\angstrom}$ Ti/$\SI{150}{\angstrom}$ Au) within $\SI{500}{\micro m}$ from the tip apex, resulting in a $\approx\SI{10}{\Omega}$ shunt resistance. 
We next cover the coarse contact pads with a thick layer of indium solder to minimize contact resistance and to improve contact with the leaf-springs used to hold the tip in the holder. We then evaporate indium in a home-built thermal evaporator at three angles to cover two contacts at $\SI{110}{\degree}$ to the apex, with a a head-on deposition performed last. The tip holder is mounted to a cryostat, and protected by radiation shielding and LN$_2$ jacketing; it is kept at a temperature of $\SI{20}{K}$ throughout the deposition process. 
Each evaporation step is preceded by 5--10 minutes of thermalization time during which the evaporator chamber is flooded with He exchange gas at a pressure of $\SI{5d-3}{mbar}$. 
Typical thicknesses for the indium depotisions are $\SI{350}{\angstrom}$ for the side contacts and $\SI{300}{\angstrom}$ for the head-on deposition for $\SI{150}{nm}$ tip diameter; we use a $\SI{1}{\angstrom/s}$ deposition rate for all steps. These parameters allow for a highly uniform, low grain-size film to form near the tip apex (see Extended Data Fig. \ref{SQUIDsensitivity}a). 

The magnetic field at the tip apex is read out by measuring the tip in a quasi-voltage bias configuration using a series SQUID array amplifier (SSAA) \cite{huber_dc_2001}. To calibrate the sensitivity of the nSOT we measure the frequency domain output of the SSAA amplifiers at the nSOT operating point. The nSOT voltage is converted to magnetic field via the transfer function, which we measure by monitoring the DC output voltage response to a $\sim\SI{20}{\micro T}$ step in magnetic field. This gives the transfer function in units of $\unit{V/T}$. As shown in Extended Data Fig. \ref{SQUIDsensitivity}b, our nSOT sensors show a maximum sensitivity (typically near a flux bias point of $\Phi_0/2$) of $\approx\SI{300}{pT/\sqrt{Hz}}$ with an effective diameter of $\SI{200}{nm}$, corresponding to $\approx \SI{6}{n\Phi_0/\sqrt{Hz}}$. 

\subsection{Magnetization reconstruction}

To reconstruct the magnetization, we follow standard Fourier domain techniques (see, e.g., the supplementary information of reference \cite{tschirhart_imaging_2021}). The magnetization shown in the main text is computed by zero-padding the measured magnetic field map before Fourier transforms are computed. This constitutes an unphysical assumption about magnetic fields outside the measurement area that were not constrained experimentally---i.e., that they vanish---and constitutes a source of systematic error.  

To address this source of error in $m$, we analyze the influence of the choice of padding on the reconstructed magnetization, shown in Extended Data Fig. \ref{OutsideSignalGapReconstruction}. Our base model assumes zero magnetic field outside the region where the fields are measured.  
We then compare this to two alternative assumptions:  first, we replicate the signal between the contacts on the left and right sides of the frame, extending the finite magnetic fields observed at the boundary into a region 500 nm wide.  As a second comparative model, we extend the fields on the edge of this region by an additional 500 nm above and below the measured region.  As shown in the figure, the reconstructed magnetization differs from the base model by $\sim10\%$ in the neighborhood of both $\nu=-1$ and $\nu=-2/3$, with  comparable or smaller effects on the inferred gap size.

\subsection{Measurement conditions for presented data}

All voltages indicated denote root-mean-square values.
\begin{itemize}
    \item Fig. \ref{fig1}a, Extended Data Figs. \ref{Devices}e, \ref{MRecon}
        \subitem Transfer function = $100$~V/T; \\height = $150~$nm;  \\$\delta V_s$ = $20$~mV; 
        \\frequency = $511.777$~Hz
    \item Fig. 1b,c, Extended Data Figs. \ref{Devices}d, \ref{Ch_displacement_field}, \ref{AbsenceOfTrivialMagnetization}
        \subitem Transfer function = $250$~V/T;  
        \\height = $100~$nm;  \\
        $\delta V_t$ = $40.1$~mV; $\delta V_b$ = $35.8$~mV; 
        \\frequency = $151.777$~Hz
    \item Fig. 1d,e
        \subitem Transfer function = $500$~V/T;  \\height = $50~$nm;  \\$\delta V_t$ = $4.1$~mV; $\delta V_b$ = $3.6$~mV; 
        \\frequency = $511.777$~Hz
    \item Fig. 1f
        \subitem Transfer function = $500$~V/T;  \\height = $50~$nm; \\ $\delta V_t$ = $40.1$~mV; $\delta V_b$ = $35.8$~mV; 
        \\frequency = $90.777$~Hz
    \item Fig. 1f inset
        \subitem Transfer function = $500$~V/T;  \\height = $50~$nm; \\ $\delta V_t$ = $20$~mV; $\delta V_b$ = $18$~mV; 
        \\frequency = $90.777$~Hz
    \item Fig. 2, Fig. 3c,d,f, Fig. 4a,b
        \subitem Transfer function = $280$~V/T;  \\height = $150~$nm;  \\$\delta V_b$ = $40$~mV;  \\frequency = $511.777$~Hz
    \item Fig. 3a,b
        \subitem Transfer function = $500$~V/T;  \\height = $100~$nm;  \\$\delta V_s$ = $5$~mV;  \\frequency = $151.777$~Hz
    \item Fig. 4d
        \subitem Transfer function = $500$~V/T;  \\height = $50~$nm;  \\$\delta V_t$ = $60.6$~mV; $\delta V_b$ = $-53$~mV; 
        \\frequency = $1151.777$~Hz
    \item Extended Data Fig. \ref{Devices}f
        \subitem Transfer function = $100$~V/T;  \\height = $180~$nm;  \\$\delta V_s$ = $28$~mV;  \\frequency = $251.777$~Hz
    \item Extended Data Fig. \ref{LLs}
        \subitem Transfer function = $500$~V/T;  \\height = $100~$nm;  \\$\delta V_b$ = $28$~mV;  \\frequency = $251.777$~Hz
\end{itemize}

\clearpage
\setcounter{figure}{0}
\renewcommand{\figurename}{\textbf{Extended Data Fig.}}
\renewcommand{\thefigure}{\arabic{figure}}

\begin{figure*}[ht!]
\includegraphics[width=4.5in]{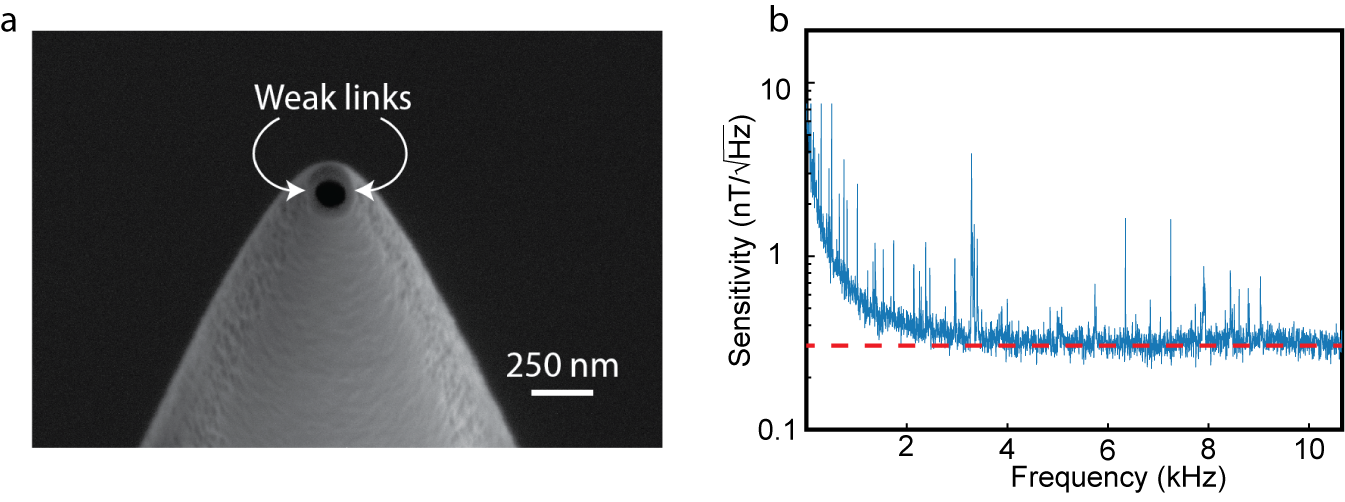}
\caption{\textbf{SQUID sensitivity}
(\textbf{a}) SEM image of the nSOT sensor used in most of the current work, highlighting the superconducting weak links on the tip apex. 
(\textbf{b}) Typical sensitivity of the nSOT as a function of the frequency. $1/f$ noise dominates at frequencies below $\SI{1}{kHz}$ and decays below the instrumentation noise floor above $\SI{2}{kHz}$ allowing for ultra-high magnetic field sensitivity well below 1nT/$\sqrt{\SI{}{Hz}}$. 
}
\label{SQUIDsensitivity}
\end{figure*}

\begin{figure*}[h]
\includegraphics[width=7in]{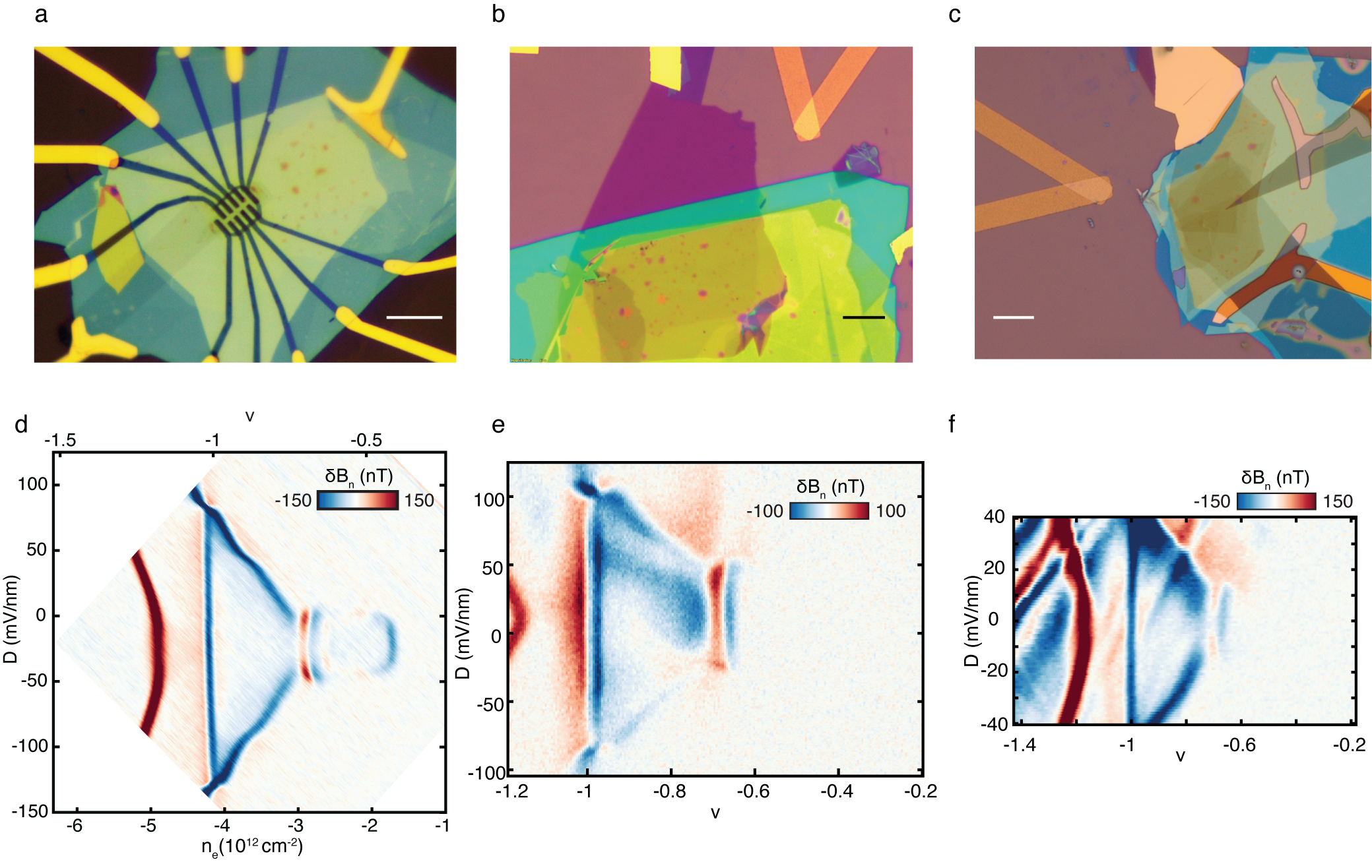}
\caption{\textbf{Devices}
(\textbf{a}) Optical micrograph of Device A (corresponding to device D(3.7$\degree$) from Ref. \cite{park_observation_2023}); 
(\textbf{b}) Device B (corresponding to device from Ref. \cite{cai_signatures_2023}); 
(\textbf{c}) Device C. Scale bar is $\SI{10}{\mu m}$; 
 (\textbf{d}) $\delta B_n$ phase diagram measured in Device A, 
 (\textbf{e}) Device B, and 
 (\textbf{f}) device C. All devices show signals near $\nu=-1$ and $\nu = -2/3$ characteristic of Chern insulators.
 }
\label{Devices}
\end{figure*}

\begin{figure*}[h]
\includegraphics[width=6.3in]{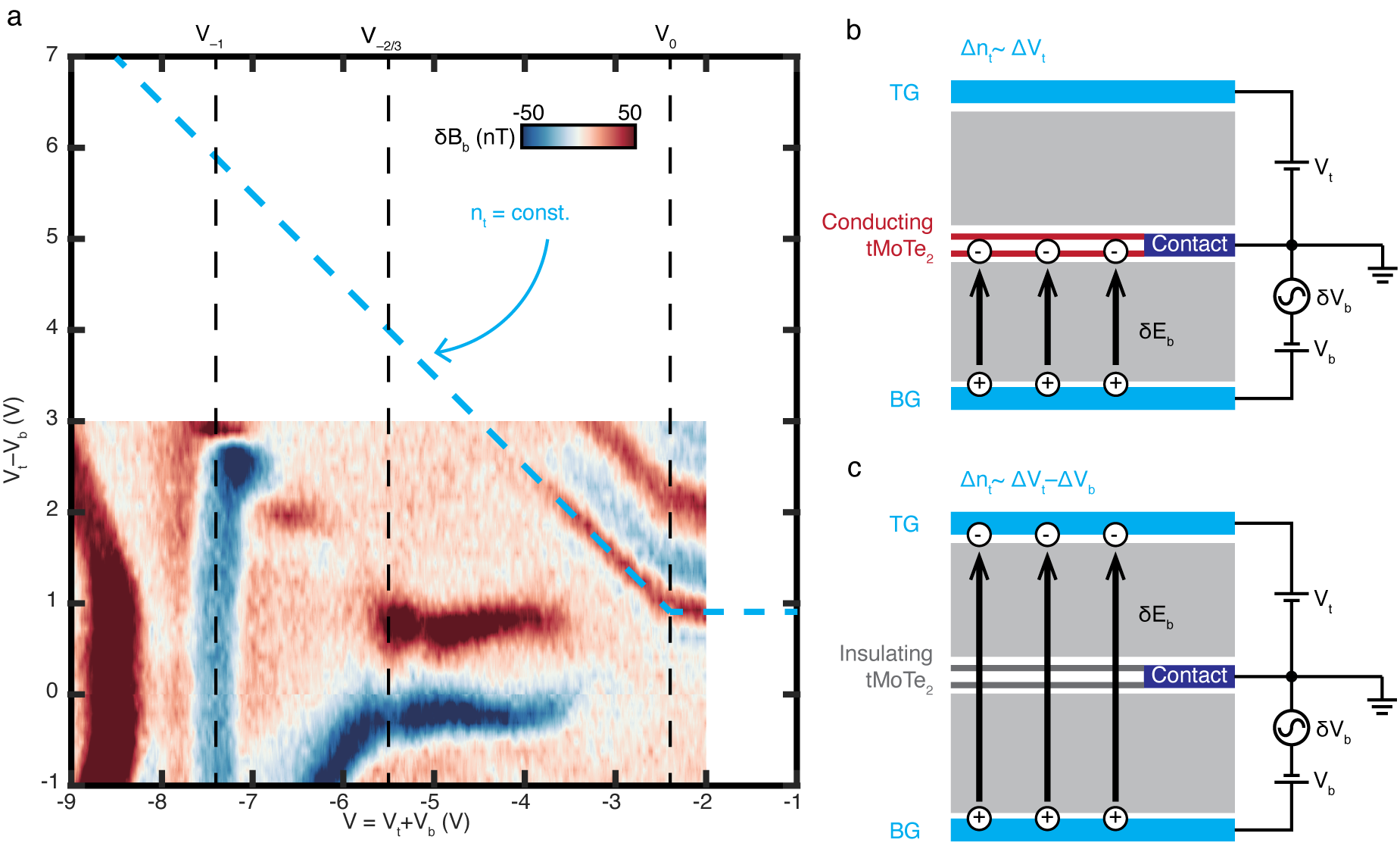}
\caption{\textbf{Chemical potential sensing and band edge offset.}
(\textbf{a}) $\delta B_b$ for device C as a function of $V\equiv V_t+V_b$ and $V_t-V_b$ exhibiting both MoTe$_2$ features described in the main text as well as features associated with Landau levels of the top graphite gate. The band edge is visible as a kink in the constant top gate carrier density trajectory, marked as a dashed blue line, with the offset $V_0$ corresponding to the voltage that separates the regime where tMoTe$_2$ is insulating and the regime where it is hole-doped. The value obtained using this method agrees with that described in the main text using the $\nu=-1$ and $\nu=-2/3$ gap densities. 
(\textbf{b}) Schematic of the electric field in the hole-doped tMoTe$_2$ regime. Here modulations of the bottom gate produce electric fields  $\delta E_b$ which are screened by the tMoTe$_2$ layer.  In this regime, the top gate density $n_t$ is tuned solely by $V_t$, and trajectories of constant $n_t$ follow slope $-1$ on the diagram in panel \textbf{a}. 
(\textbf{c}) Schematic of the electric field in the insulating tMoTe$_2$ regime.  Here $\delta E_b$ penetrates the tMoTe$_2$, so that $n_t$ is tuned by both $V_t$ and $V_b$. In this regime, constant-$n_t$ lines have slope $0$ in panel \textbf{a}. 
}
\label{LLs}
\end{figure*}

\begin{figure*}[ht!]
\includegraphics[width=4.5in]{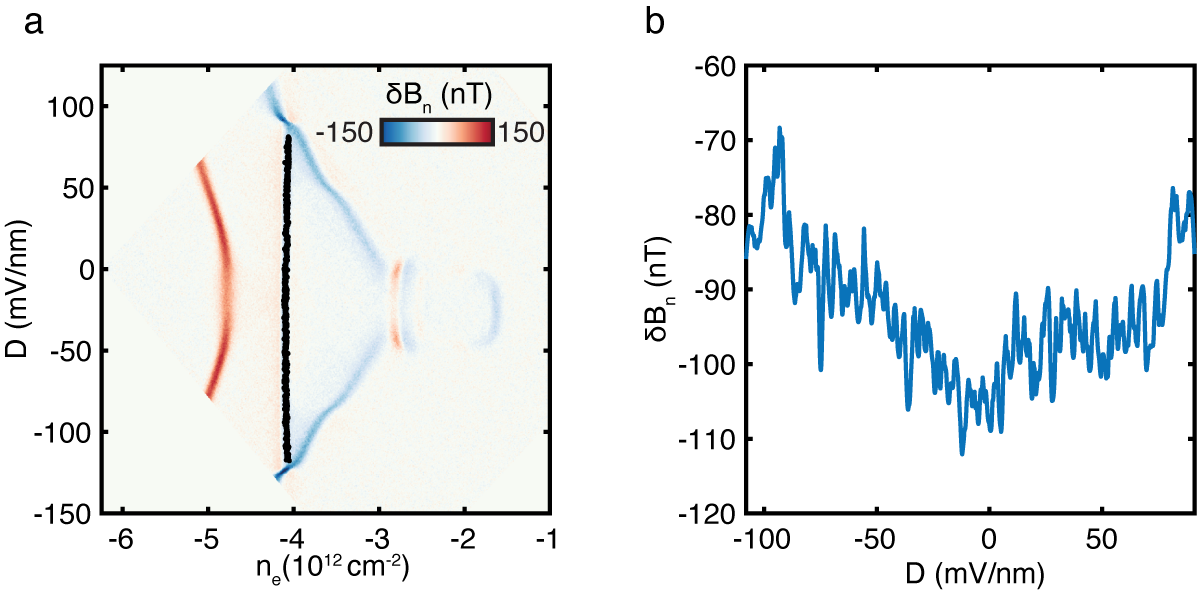}
\caption{\textbf{Chern insulator signal dependence on the displacement field}
(\textbf{a}) Signal $\delta B_n$ as a function of charge carrier density and electric displacement field;
(\textbf{b}) Minimum value of $\delta B_n$ in a window around $\nu = -1$ gap along displacement field axis. The positions of the values are shown on \textbf{a}. The signal from the gap decays slightly as a function of displacement field, remaining finite up to the first order phase transition induced by layer polarization.  
}
\label{Ch_displacement_field}
\end{figure*}

\begin{figure*}[ht!]
\includegraphics[width=5.2in]{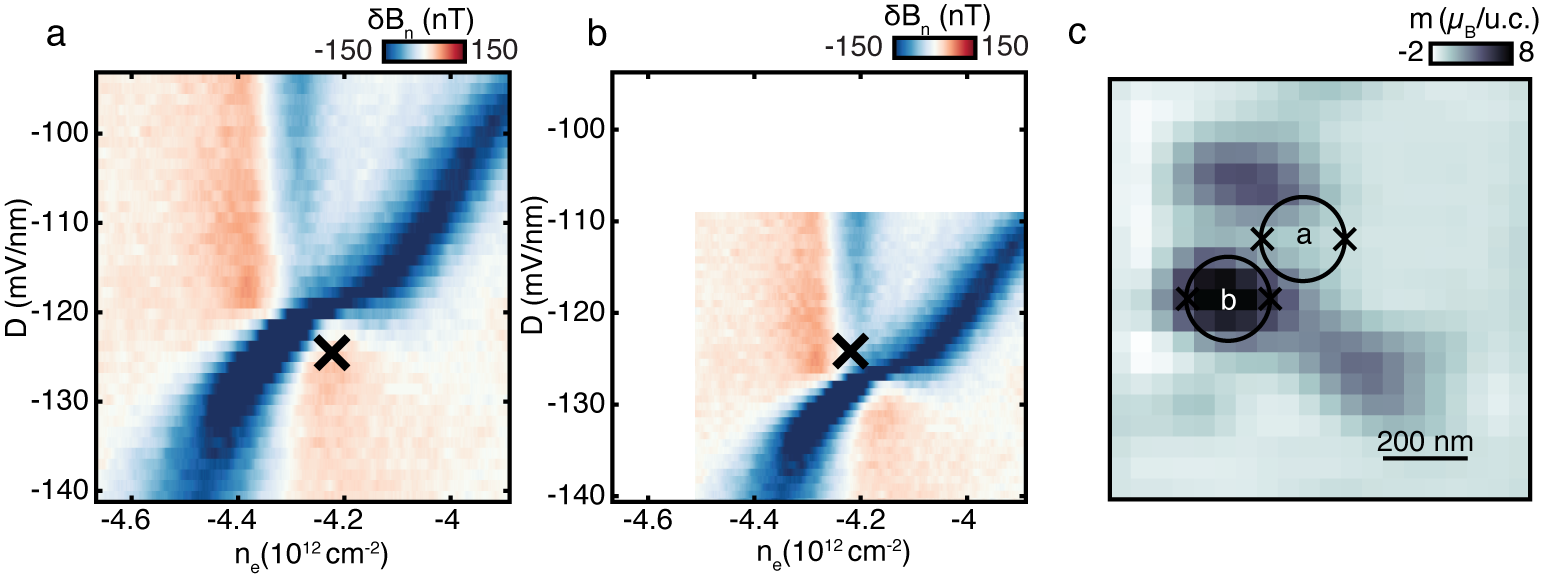}
\caption{\textbf{First order valley polarization transition at $\nu=-1$} 
(\textbf{a}) $\delta B_n$ near the valley polarization transition in the vicinity of $\nu=-1$ at point ``a'' in panel c. 
(\textbf{b}) $\delta B_n$ measured at point ``b'', \SI{200}{nm} away from point ``a''.  
Both panel a and b show a sharp first-order-like signal on the boundary of the $\nu=-1$ Chern insulator peak, but these transitions appear at slightly different values of $D$ and $n_e$.
(\textbf{c}) Reconstructed magnetization in the point indicated by the ``$\times$'' on panels a and b.  The measured measured $m$ at position ``a'' is consistent with zero to within our experimental error, and we find no evidence for non-zero net magnetization in the high $|D|$ phases. The non-zero signal observed in $\delta B_n$ at that position is associated with fringe AC magnetic fields arising from areas where where the ``$\times$'' position in parameter space corresponds to the valley transition. 
}
\label{AbsenceOfTrivialMagnetization}
\end{figure*}

\begin{figure*}[ht!]
\includegraphics[width=5.2in]{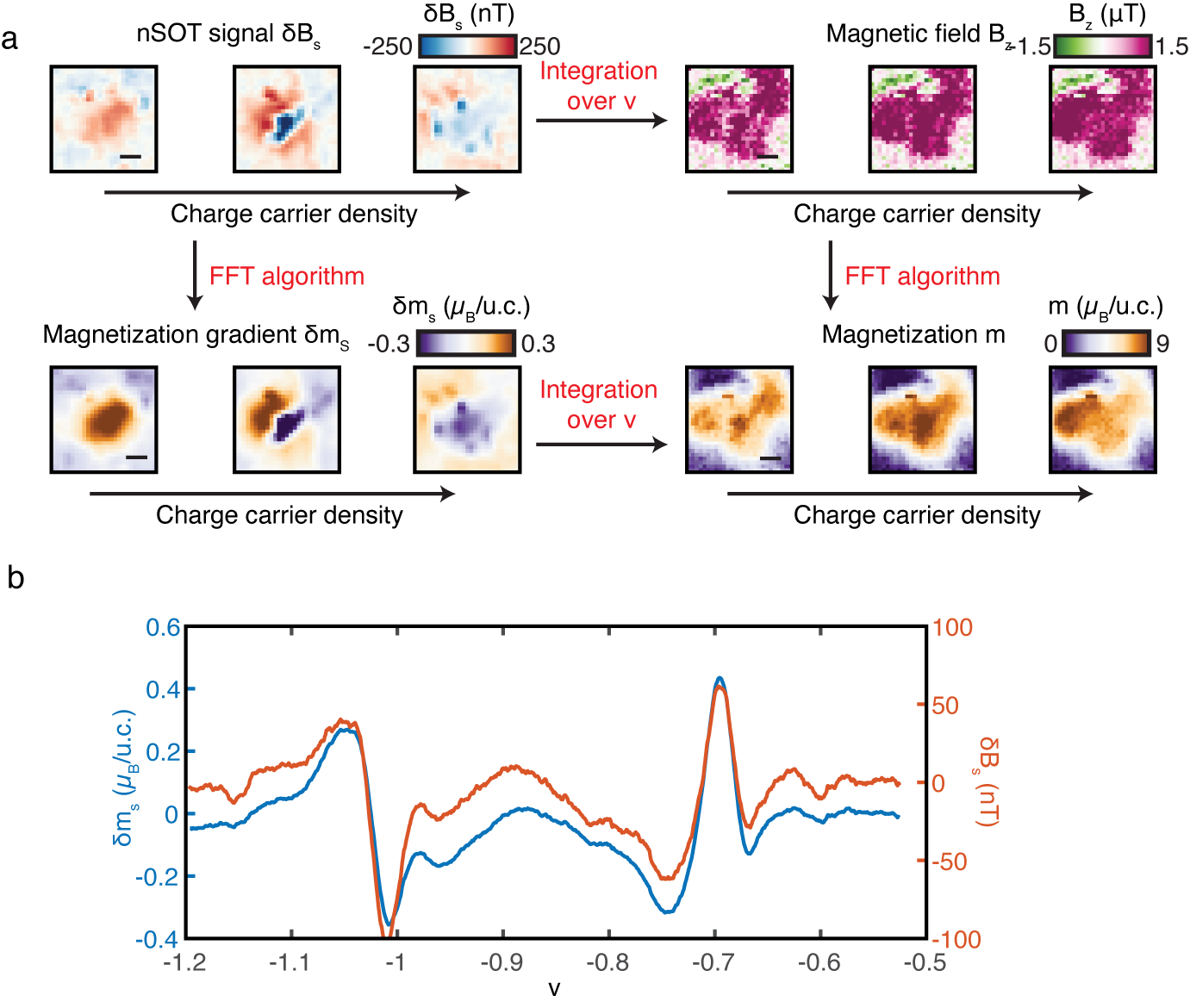}
\caption{\textbf{Magnetization reconstruction.}
(\textbf{a}) Schematic of analysis to reconstruct the magnetization. We measure the local magnetic field $\delta B_s$ in response to a modulated sample voltage $\delta V_s$; data is from Device B, with scale bar of 2~$\mu$m. An FFT-based algorithm (see Methods) can be used to directly compute the corresponding $\delta m_s$, which may then be integrated over $V_s$ to obtain m.  Alternatively, the $\delta B_s$ signal may be integrated over $V_s$ to obtain $B$, which can in some cases be compared to the directly measured $B_{DC}$; this can them be processed by the FFT algorithm to produce the same $m$.   
(\textbf{b}) Comparison of the measured $\delta B_s$ (orange) and the reconstructed $\delta m_s$ (blue) as a function of the filling factor $\nu$ at a single given spatial location. While qualitative features are preserved, but of course quantitative features differ. }
\label{MRecon}
\end{figure*}

\begin{figure*}[ht!]
\includegraphics[width=6.6in]{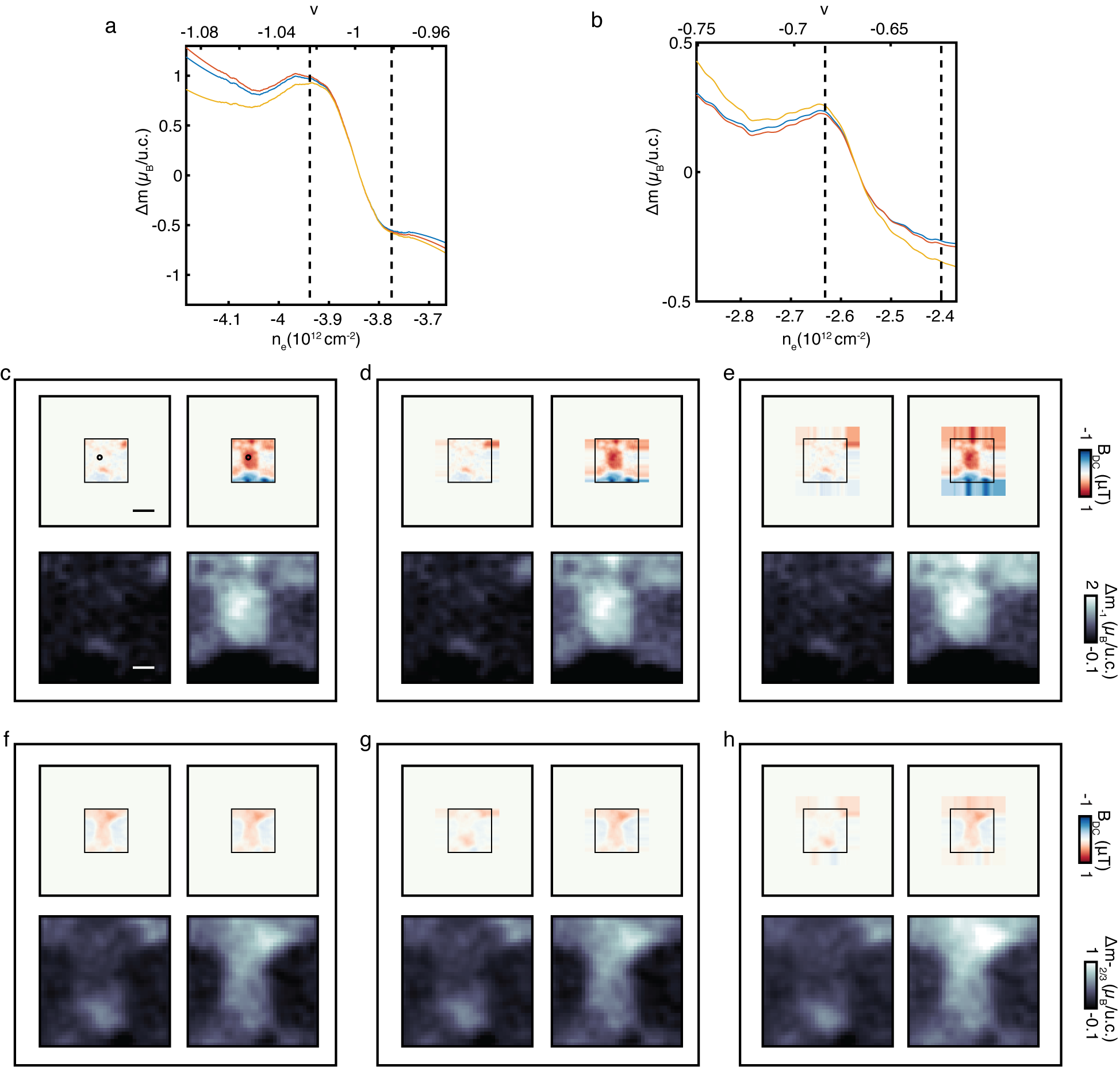}
\caption{
\textbf{Estimating systematic error due to padding assumptions.}
(\textbf{a}) Magnetization change $\Delta m$ (referenced to the center of the $\nu=-1$ gap) in the same location as that shown in Fig. \ref{fig3} for the three different padding assumptions described in panels c, d, and e.  
(\textbf{b}) Magnetization change $\Delta m$ (referenced to the center of the $\nu=-2/3$ gap) in the same location as that shown in Fig. \ref{fig3} for the three different padding assumptions described in panels f, g, and h.
(\textbf{c}) Top row: measured $B_{DC}$ with zero padding assumption across the $\nu=-1$ gap. Scale bar is \SI{600}{nm}. Left and right panels correspond to $n_e$ values shown by dotted lines in panel a.  
Bottom row: reconstructed magnetization in the measured range at the same positions, with the zero padding assumption.  The magnetization change across the gap is \SI{1.44}{\mu_B}/u.c. Scale bar is \SI{200}{nm}. 
(\textbf{d}) Same as panel c, except with the padding assumption that data extends to the right and left of measured area as shown. The magnetization change across the gap is \SI{1.48}{\mu_B}/u.c. 
(\textbf{e})Same as panels c and d but with additional padding as shown. Magnetization change across the gap is \SI{1.41}{\mu_B}/u.c.
(\textbf{f}) Top row: measured $B_{DC}$ with zero padding assumption across the $\nu=-2/3$ gap.
 Left and right panels correspond to $n_e$ values shown by dotted lines in panel \textbf{b}.  
Bottom row: reconstructed magnetization in the measured range at the same positions, with the zero padding assumption.  The magnetization change across the gap is \SI{0.46}{\mu_B}/u.c. 
(\textbf{g}) Same as panel f, with the same padding assumption as panel \textbf{d}. Magnetization change across the gap is \SI{0.46}{\mu_B}/u.c.; 
(\textbf{h}) Same as panels f-g but with the same padding assumption as in panel\textbf{e}. Magnetization change across the gap is \SI{0.51}{\mu_B}/u.c.}
\label{OutsideSignalGapReconstruction}
\end{figure*}

\end{document}